\documentclass[reprint, amsmath,amssymb, aps]{revtex4-2}
\usepackage{amssymb}
\usepackage{bm}
\usepackage{url}
\usepackage{graphicx}
\usepackage{color}

\begin{document}

\title{Turbulence dictates the fate of virus-containing droplets in violent expiratory events}

\author{M.~E. Rosti$^{1}$}
\author{M. Cavaiola$^{2,3}$}
\author{S. Olivieri$^{1}$}
\author{A. Seminara$^{4,5}$}
\author{A. Mazzino$^{2,3}$}
\email[Corresponding author:]{andrea.mazzino@unige.it}

\affiliation{
$^1$ Complex Fluids and Flows Unit, Okinawa Institute of Science and Technology Graduate University, 1919-1 Tancha, Onna-son, Okinawa 904-0495, Japan\\ 
$^2$ Department of Civil, Chemical and Environmental Engineering (DICCA), University of Genova, Via Montallegro 1, 16145, Genova, Italy \\
$^3$ INFN, Genova Section, Via Montallegro 1, 16145, Genova, Italy \\
$^4$ CNRS, Institut de Physique de Nice, UMR7010, 06108 Nice, France\\
$^5$ Universit\'e C\^ote d’Azur, Institut de Physique de Nice, UMR 7010, 06108 Nice, France
}

\begin{abstract}
Violent expiratory events, such as coughing and sneezing, are highly nontrivial examples of a
two-phase mixture of liquid droplets dispersed into an unsteady turbulent airflow.
Understanding the physical mechanisms determining the dispersion and evaporation process of respiratory droplets has recently become a priority given the global emergency caused by the SARS-CoV-2 infection.
By means of high-resolution direct numerical simulations (DNS) of the expiratory airflow and a comprehensive Lagrangian model for the droplet dynamics, we identify the key role of turbulence on the fate of exhaled droplets. Due to the considerable spread in the initial droplet size, we show that the droplet evaporation time is controlled by the combined effect of turbulence and droplet inertia.
This mechanism is clearly highlighted when comparing the DNS results with those obtained using coarse-grained descriptions that are employed in the majority of the current state-of-the-art investigations, resulting in errors up to $100\%$ when the turbulent fluctuations are filtered or completely averaged out.
\end{abstract}


\maketitle 


\section{Introduction}
Turbulent transport of droplets in a jet/puff is a problem of paramount importance in science and engineering  that nowadays has become even more important given the global emergency caused by the COVID-19 infection; for a recent review see e.g.~\cite{lincei2020review,mittal2020flow,bahl2020airborne}. The relationship stems from the fact that
the dominant route of SARS-CoV-2 spread is via small virus-containing respiratory droplets that the infected person exhales when coughing, sneezing or talking~\cite{asadi2019aerosol}.
The spread thus does not necessarily involve a physical contact between the infected and the susceptible persons \cite{lewis2020coronavirus}.

\begin{figure*}
\centering
\includegraphics[width=0.95\textwidth]{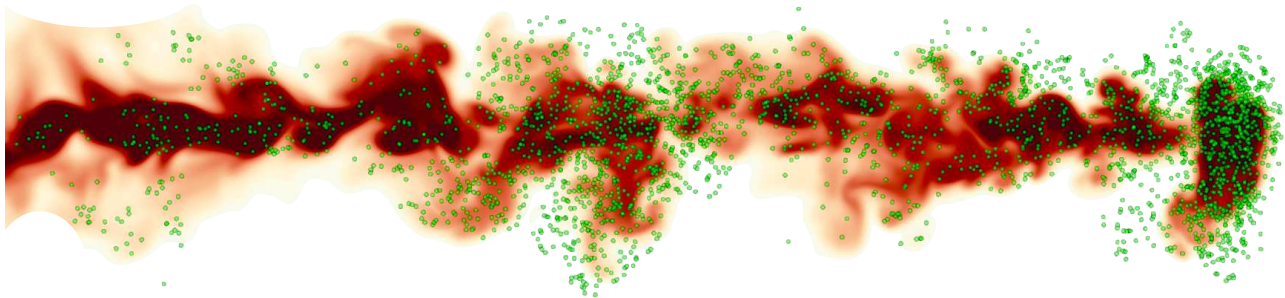}  
\caption{Snapshot of the expiratory event 7.6 s after start coughing obtained from our numerical simulations. Different colors represent different values of the humidity field ranging between the 99$\%$ of the ambient humidity $\textrm{RH}_a$ (red areas) and $\textrm{RH}_a$ (white areas). Green bullets (shown not in scale) identify the position of the airborne  droplets, initialised with the sizes taken from~\citet{duguid1946size}. The streamwise extension of the puff at this time is 2.6 m.}
\label{fig:sketch}
\end{figure*}

Because the exhalation process occurring in violent air expulsions (e.g., coughing and sneezing) has a finite duration, one has to distinguish between two different regimes for the evolution of the exhaled airflow (a cloud in short): one related to the early evolution stage and one related to the late evolution stage.
In the initial stage of the evolution, which defines the jet phase~\cite{celani2014odor,ghaem2010velocity}, the mouth (i.e., the source) is still injecting air into the ambient; in the late stage, which defines the puff phase~\cite{ghaem2010velocity,kovasznay1975unsteady}, the cloud stops to receive momentum from the source and becomes freely evolving in the ambient. The initial jet behavior is determined by the conservation of the momentum flux $\rho r^2 (dx/dt)^2 \sim constant$ together with the assumption of a self-similar behavior (i.e.,~a power law) for the longitudinal coordinate of the cloud center of mass $x(t)$ and the cloud radius $r(t)\propto x(t)$. Combining these ingredients, one easily gets $x(t)\sim t^{1/2}$ for the jet phase~\cite{bourouiba2014violent}. Note that in violent human expulsions (e.g.,~for a cough) the momentum flux at the source is time dependent \cite{gupta2009flow}, a fact that may cause the breakdown of the self-similarity hypothesis for the jet phase.

In the puff stage, the momentum of the cloud is constant
$\rho r^3 (dx/dt) \sim constant$ which, again under the hypothesis of self-similarity and $r(t)\propto x(t)$, leads to $x(t)\sim t^{1/4}$ \cite{bourouiba2014violent,kovasznay1975unsteady}. Owing to the chaotic/turbulent  nature of the jet phase, the puff  behavior is expected to be robust with respect to different ways of producing the air expulsion at the source.

Jet  and puff  phases, because of their different  self-similar behaviors, are thus expected to  affect  in a different way the transport process of droplets hosted in the flow.
The transport process of momentum is indeed further complicated by the fact that the released cloud may hardly be interpreted as a homogeneous fluid. It rather consists of a two-phase mixture of droplets dispersed into a fluid phase which is hotter and more humid than the ambient air. Two new players are thus involved in the transport process: the humidity field
(or, equivalently, the  supersaturation field for almost isothermal expulsion processes) and
droplet evaporation.

There is however one additional player in violent respiratory events: fluid turbulence. The typical duration of a cough is 200-500 ms, the average mouth opening of male subjects is (4 $\pm$ 0.95) $\mathrm{cm^2}$, and the resulting Reynolds number is about $10^4$~\cite{gupta2009flow,bourouiba2014violent}.  Larger values for the Reynolds number (of about a factor 4) have been found for a sneeze expulsion~\cite{bourouiba2014violent}. One more player can thus contribute to  dictate the fate of ejected droplets and thus of the virus spreading: turbulent fluctuations of both the carrier flow and of the humidity field.

Understanding the combined role of turbulence and droplet inertia on the virus-containing droplet evaporation under realistic conditions mimicking a human cough is the main aim of the present work.  To do that, we attack the problem on the numerical side by performing accurate Direct Numerical Simulations (DNS) for the fluid flow and humidity field,
complemented by a Lagrangian solver for the droplet dynamics including a  dynamical equation for the evolution of the droplet radii modeling the evaporation/condensation process (see Fig.~\ref{fig:sketch}).
Such an accurate description is nowadays possible thanks to the deep understanding achieved in the microphysics of small liquid droplets under different ambient conditions \cite{pruppacher1997microphysics}.

For the problem of respiratory droplet spreading, typical approaches found in the current literature are based on Large-Eddy Simulations (LES) and Reynolds-Averaged Navier--Stokes equations (RANS) (see e.g., \cite{zhao2005numerical,zhu2006transport,chen2010some,li2018modelling,feng2020influence}).
By definition, LES and RANS only describe turbulent fluctuations at the largest scales involved. On the other hand, the fine structure of turbulence is expected to be crucial to correctly account for its effect on droplet evaporation. This is expected from results in atmospheric cloud
microphysics where turbulence is crucial to explain the broadening of the cloud-droplet size spectrum  (see e.g.~\citet{celani2005droplet,celani2008equivalent,celani2009droplet}).
Numerical approaches based on DNS are thus crucial to assess quantitatively how turbulence dictates the fate of virus-containing
droplets, and consequently provide useful insights on the spread of SARS-CoV-2 and other airborne transmitted infections.

The rest of the paper is organized as follows: in Sec.~\ref{sec:method} we introduce the methodology of the investigation, in Sec.~\ref{sec:results} we compare droplet fate in simulations that ignore or simplify turbulence vs simulations that fully account for turbulence. Finally, in Sec.~\ref{sec:conclusions} we draw the concluding remarks.

\vspace{\fill}

\section{Method}
\label{sec:method}

\subsection{Governing equations}

The airflow exhaled from the mouth is ruled by the incompressible Navier--Stokes equations
\begin{equation}
  \partial_t \bm{u} + \bm{u}\cdot \bm{\partial} \bm{u}=-\frac{1}{\rho_a}\bm{\partial}p + \nu\partial^2 \bm{u} \qquad \bm{\partial}\cdot \bm{u}=0
  \label{eq:NS}
  \end{equation}
with $\nu$ being the air kinematic viscosity and $\rho_a$ the air density. The list of all relevant parameters used in this study are reported in Appendix~\ref{app:properties}. 
Instead of simulating the evolution of the absolute humidity field (the exhaled air is saturated, or close to saturation \cite{morawska2009size}) it is more convenient to model directly the supersaturation field (i.e. $s=\mathrm{RH}-1$, $\mathrm{RH}$ being the relative humidity).
Indeed, the supersaturation dictates the evaporation/condensation process, as it appears in the evolution equation for droplet radius  \cite{pruppacher1997microphysics}.
The supersaturation field is ruled by the  advection-diffusion equation \cite{celani2005droplet}:
\begin{equation}
  \partial_t s + \bm{u}\cdot \bm{\partial} s= D_v \partial^2  s,
\label{eq-supersat}
\end{equation}
$D_v$ being the water vapor diffusivity.
Eq.~\eqref{eq-supersat} assumes that the saturated vapor pressure is constant, an assumption that holds as long as the ambient is not much colder than the exhaled air, which is at about $30\,\mathrm{^oC}$ according to~\citet{morawska2009size}.

\begin{figure}
    \centering
    \includegraphics[width=0.45\textwidth]{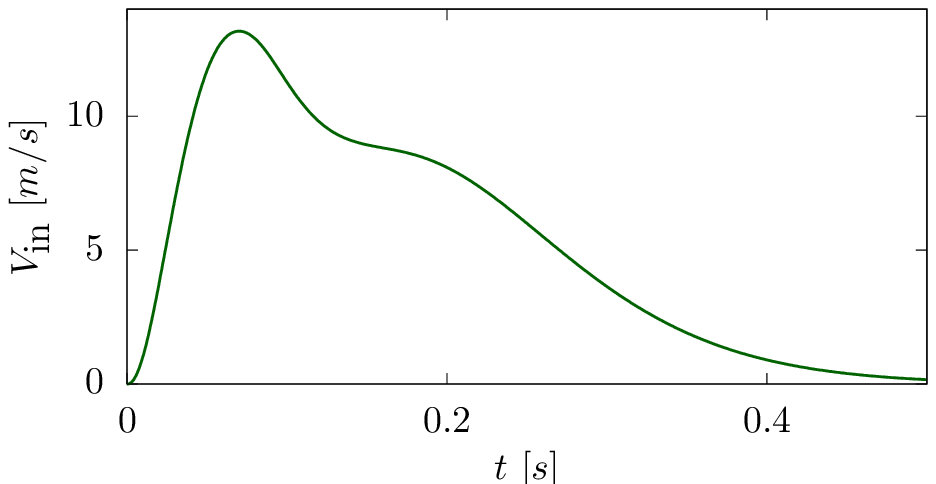}\\[11pt]
    \includegraphics[width=0.45\textwidth]{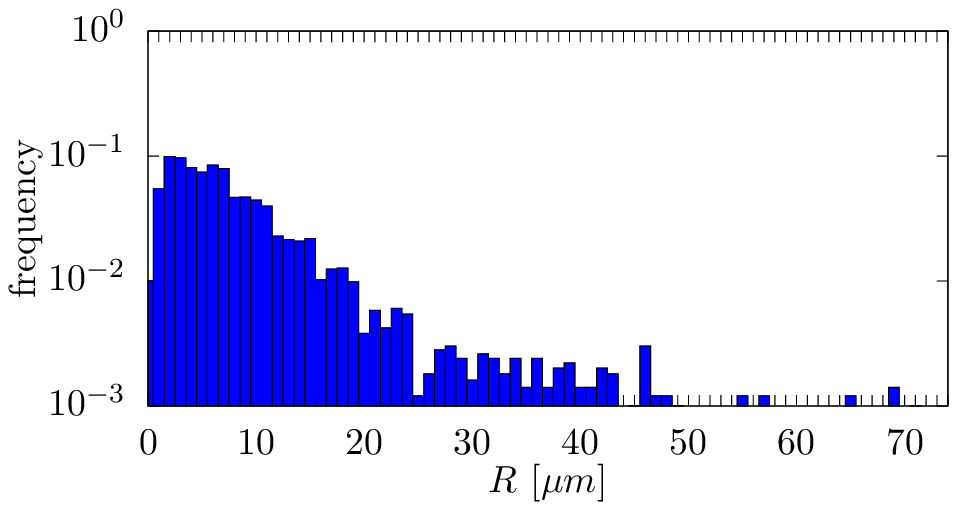}
    \caption{(top) Time-varying inlet air velocity representative of cough according to Ref.~\cite{gupta2009flow}. (bottom) Droplet initial size distribution according to Ref.~\cite{duguid1946size}.}
    \label{fig:inletVel}
\end{figure}

To simulate the airflow generated by human cough, we adopt the inlet air velocity profile proposed by~\citet{gupta2009flow}, as shown in Fig.~\ref{fig:inletVel} (top). The air is assumed to be saturated (i.e., $s=0$) as it exits from the mouth opening of area 4.5 $\mathrm{cm^2}$. The duration of the expulsion  is approximately 0.4 s and the peak velocity is 13 m/s. The resulting Reynolds number (based on the peak velocity and on the mouth average radius) is about 9000. The flow field is thus fully turbulent as one can easily realize by looking at Fig.~\ref{fig:sketch}.

\begin{figure}
\centering
\includegraphics[width=0.49\textwidth]{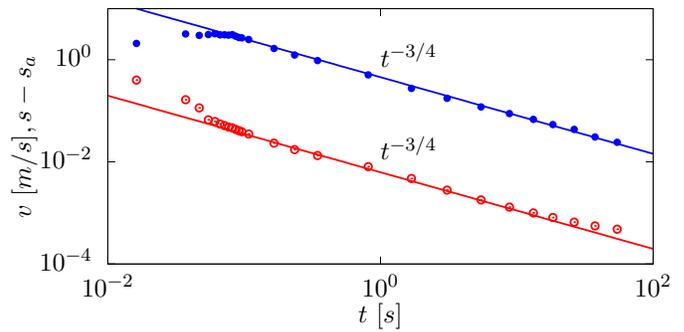} 
\caption{Mean velocity $v$ of the exhaled air (blue, filled circles) and of the supersaturation $s-s_a$ (red, open circles) as a function of time.
Blue (red) lines show the scaling predicted for the velocity field in  Ref.~\cite{kovasznay1975unsteady} which also holds for the supersaturation field.}
\label{fig:puff}
\end{figure}

Before discussing how the liquid part of the two-phase mixture is modelled, let us first validate the puff dynamics of the exhaled air. 
By means of a simple phenomenological approach, we show how one can derive  the temporal scaling for the standard deviation of a cloud of tracers in a turbulent puff.  The starting point is the result obtained by~\citet{kovasznay1975unsteady} for the temporal scaling of the puff radius $\sigma^u \sim t^{1/4}$, obtained by the authors in terms of a simple eddy-viscosity approach. In order to determine the standard deviation, $\sigma$, for a cloud of tracers carried by the turbulent puff, one has to resort to the concept of relative dispersion. The latter can be described in terms of arguments {\it \`a la} Richardson~\cite{richardson1926atmospheric}. Accordingly, $\sigma (t) \sim \epsilon(t)^{1/2} t^{3/2}$, where $\epsilon(t)$ is the turbulence dissipation rate. This latter can be easily estimated from the well-known $4/5$th Kolmogorov law evaluated at the integral scale $\sigma^u$. Namely,
\begin{equation}
  \epsilon (t) \sim \frac{\delta U^3}{\sigma^u}\quad \mathrm{with} \quad \delta U \sim \frac{\sigma^u}{t}
  \label{eq:epsilon}
  \end{equation}
from which one immediately gets: $\epsilon (t) \sim t^{-5/2}$.
The scaling law for $\epsilon$  immediately leads to the temporal scaling for the standard
deviation of the tracer cloud: $\sigma (t) \sim t^{1/4}$.
Finally, because $\langle s \rangle $ is proportional to the puff volume, and this latter goes as $\sigma^3\sim t^{3/4}$,  the decay law for the mean supersaturation is: $\langle s(t) \rangle \sim t^{-3/4}$.
The same law holds for the mean streamwise puff velocity~\cite{kovasznay1975unsteady}.
The reliability of our puff dynamics
is demonstrated in Fig.~\ref{fig:puff} which clearly shows the expected scaling laws for more than two decades with high accuracy.

We are now ready to introduce the model for the liquid part of the two-phase mixture.
It is described as an ensemble of $N$ inertial particles ruled by the well-known set of equations \cite{maxey1983equation}:
\begin{equation}
  \dot{\bm{X}_i}=\bm{U}_i(t)+\sqrt{2 D_v}\bm{\eta}_i (t)\qquad i=1, \ldots , N
  \label{eq:max1}
  \end{equation}
  \begin{equation}
  \dot{\bm{U}_i}=\frac{\bm{u}(\bm{X}_i(t),t)-\bm{U}_i(t)}{\tau_i} +\bm{g}
\label{eq:max2}
\end{equation}
\begin{equation}
\tau_i =\frac{2 (\rho_{D\,i}/\rho_a) R_i^2(t)}{9\nu}
\label{eq:tau}
\end{equation}
where $N$ is the number of exhaled droplets (here N$\approx$ 5000 according to~\citet{duguid1946size}),
$\bm{X}_i$ is the position of the i-th droplet and $\bm{U}_i$ its velocity, and, finally, $\bm{g}$ is the gravitational acceleration.
Each droplet is affected by a Brownian  contribution via the white-noise process $\bm{\eta}_i$ (see also Appendix~\ref{app:droplet}).
Here, $\rho_{D\,i}$ is the density  of the i-th droplet.
Finally, $\tau_i $ is the Stokes relaxation time of the  $i$-th droplet and $R_i$ is its radius.

Since in our case the flow is neither statistically homogeneous nor stationary, we consider the characteristic flow time scale $\tau_\mathrm{flow} = \sqrt{\nu \sigma_u/v^3}$
where $v$ is the puff mean velocity measured by the Lagrangian tracers (as later described in Sec.~\ref{app:coarse-graining}).
Using the latter, we can define the Stokes number for the $i$-th droplet as $\mathit{St}  = {\tau_i}/{\tau_\mathrm{flow}}$,
which allows us to clearly distinguish droplets whose trajectory is (or is not) dominated by inertia, i.e. $\mathit{St} > 1$ (or $\mathit{St} < 1$).

 Droplets are assumed to be made of salt water (water and NaCl) and a  solid insoluble part (mucus) \cite{vejerano2018physico}\footnote{The assumption corresponds to their model 1 for respiratory fluids reported in their Table 1}.  
Droplet radius evolves according to the ruling equation~\cite{pruppacher1997microphysics}
\begin{equation}
  \frac{\mathrm{d}}{\mathrm{d}t} R^2_i(t) = 2 C_R\left (1+s(\bm{X}_i(t),t)-e^{\frac{A}{R_i(t)}-B\frac{r_{N\,i}^3}{R_i^3(t)-r_{N\,i}^3 } }       \right )
\label{eq-radii}
\end{equation}
\begin{equation}
  R_i(t)=r_{N\,i}  \qquad \mbox{for}\qquad s\le s_{crh} \; \mbox{(crystallization)}
  \label{eq:R}
\end{equation}
 No feedback of this equation to Eq.~(\ref{eq-supersat}) is considered here because of the very small values of the liquid volume fraction, typically smaller than $10^{-5}$~\cite{wang1993settling,bourouiba2014violent} or even smaller according to~\citet{johnson2011modality} and~\citet{morawska2009size}, and thus droplet back reaction  on the flow is largely negligible.
In Eq.~(\ref{eq-radii}), $C_R$ is the droplet condensational growth rate,
$s_{crh}=-0.55$  ($CRH=0.45 $, the  so-called crystallization RH or efflorescence RH)  for NaCl \cite{lohmann2016introduction}. 
Fig.~3 of Ref.~\cite{zeng2014temperature} and Ref.~\cite{biskos2006nanosize} show
the weak dependence of CRH on temperature. $r_{N\,i} $ is the radius of the (dry) solid part of the i-th droplet when the salt is entirely crystallized (i.e. below CRH). 
The dependence  of $r_{N\,i} $ on physical/chemical/geometrical properties of the exhaled droplets is reported in Appendix~\ref{app:properties} together with the expressions of parameters $A$ and $B$. On the basis of the parameters assumed here, the ratio $r_{N\,i}/R_i(0)$ is 0.16
 which agrees with the estimations discussed in~\citet{nicas2005toward}.
 
We consider here the initial distribution of droplet sizes to match seminal experiments by~\citet{duguid1946size}, which is still considered as a reference on the subject.
According to~\citet{duguid1946size} and as shown in Fig.~\ref{fig:inletVel} (bottom), we consider initial droplet radii approximately ranging from $1$ to $1000\, \mathrm{\mu m}$ with the $95\%$ falling between 1 and $50\, \mathrm{\mu m}$.
Droplets are initially at rest and randomly distributed within a sphere of radius $1\,\mathrm{cm}$ located inside a pipe conceptually mimicking the human mouth (see Appendix~\ref{app:num-meth}).
Finally, the exhaled droplets enter the ambient considered initially at rest with a relative humidity $\mathrm{RH}_a=60 \%$ (i.e. $s_a=-0.4$), larger than the crystallization RH.
Note that states of local equilibrium are possible from Eq.~\eqref{eq:R} owing to the solute effect~\cite{pruppacher1997microphysics}.

\subsection{Numerical method}
The employed in-house flow solver is named \textit{Fujin} (\texttt{https://groups.oist.jp/cffu/code}) and is based on the (second-order) central finite-difference method for the spatial discretization and the (second-order) Adams-Bashfort scheme for the temporal discretization. The Poisson equation for the pressure is solved using the \texttt{2decomp} library coupled with a fast and efficient FFT-based approach.   The solver is parallelized using the MPI protocol and has been extensively validated in a variety of problems~\cite{rosti_brandt_2017a,rosti2019flowing,rosti_ge_jain_dodd_brandt_2019,rosti2020increase,olivieri2020dispersed}. The droplet dynamics is computed via Lagrangian particle tracking complemented by an established droplet condensation model that has been successfully employed in the past for the analysis of rain formation processes~\cite{celani2005droplet,celani2008equivalent,celani2009droplet}. The governing equations for the droplet dynamics [Eqs. \eqref{eq:max1}--\eqref{eq:R})] are advanced in time using the explicit Euler scheme. The numerical domain is discretized with a uniform grid of size $3.5\,\mathrm{mm}$ and we verified that the following results are independent of the grid size, statistical sample and droplet initial condition.
Further information on the numerical setup, method and verifications are reported in Appendix~\ref{app:num-meth}.
 
 \subsection{Coarse-graining approaches}
\label{app:coarse-graining}

In this work, we aim at highlighting the crucial role of turbulence on the dynamics of expiratory droplets. To this aim, two additional types of coarse-grained simulations have been performed as detailed in the following of this section.

\subsubsection{Filtered DNS}
In the so-called filtered DNS, we let the governing equations, i.e. the Navier-Stokes equations for the fluid flow and the advection-diffusion equation for the supersaturation field, evolve exactly as in the fully-resolved DNS. However, both in the Lagrangian particle tracking and in the droplet radii evolution equation [Eqs. \eqref{eq:max1}--\eqref{eq:R})], instead of using the actual fluid velocity/supersaturation, we make use of their averaged values over a stencil of $7^3$ Eulerian grid points surrounding the droplet. As a result, the fine structure of both the velocity and supersaturation fields is washed out.

\subsubsection{Mean-field simulation}
In this last approach, we first seed the fluid flow with $20000$ Lagrangian  tracers from which we reconstruct a mean, time-dependent streamwise velocity field (whereas both the spanwise and the vertical components are set to zero because of the symmetry of the problem) and a mean, time-dependent supersaturation field of the turbulent puff. Such mean velocity is thus supplied to the Lagrangian particle tracking while the mean supersaturation field is supplied to the droplet radii evolution equation [Eqs. \eqref{eq:max1}--\eqref{eq:R})]. Moreover, from the tracer trajectories we also measure the time evolution of the puff size.  The latter is used to specify at each iteration whether the droplet resides inside or outside the puff. In the first case, we apply the described mean fields; conversely, outside the puff we impose $s=s_a$ and $\bm{u}=\bm{0}$.

\section{Results}
\label{sec:results}

\begin{figure}
\centering
\includegraphics[width=0.49\textwidth]{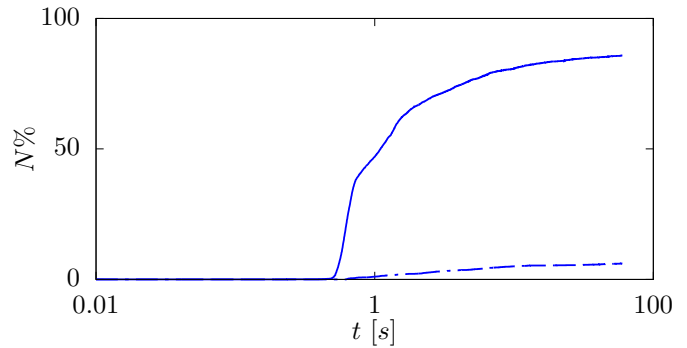} 
\caption{Time history of the percentage number of droplets settling on the ground (dashed line) vs those remaining airborne and reaching $1\,m$ from the mouth (solid line).}
\label{fig:Nt}
\end{figure}

In this work, we focus on airborne transmitted droplets where turbulence is expected to play a significant role. Nevertheless, as a first step in our analysis, we provide an overview of the observed dynamics by quantifying the number of airborne transmitted droplets and of those settling on the ground.  Such information is reported in Fig.~\ref{fig:Nt}, from which we clearly observe that the number of sedimenting droplets represents only a tiny fraction (around $5\%$) of the total number of exhaled droplets.
Sedimenting droplets have larger size and are characterized by a ballistic-like trajectory, due to the fact that the effect of gravity largely dominates the action exerted by the flow. Because the dynamics of these droplets is ballistic, we do not further discuss their fate but rather focus on the behavior of airborne transmitted droplets, with a particular attention to the role of turbulent fluctuations both in their dispersion and evaporation process.

Because the supersaturation field evolves as a passive scalar in a turbulent field, it exhibits the well known ``plateaux-and-cliffs'' structures
\cite{shraiman2000scalar,falkovich2001particles,warhaft2000passive,watanabe2004statistics,celani2001fronts}. Namely, the scalar field displays dramatic fluctuations occurring in small regions (called cliffs or fronts) separating larger areas where the scalar is well mixed (called plateaux).  Because airborne droplets and supersaturation are transported by the same velocity field, correlations occur between droplet trajectories and supersaturation values~\cite{celani2005droplet}. 
This phenomenon causes droplets of sufficiently small size to remain long in the large well-mixed regions where they can equilibrate with the (local) value of the supersaturation.  The droplet evaporation process is thus expected to behave in time by alternating phases of  equilibrium with phases of rapid evaporation, i.e., a sort of stop-and-go process. The same type of structures is also expected for the decay of droplet radii.
\begin{figure}
\centering
\includegraphics[width=0.49\textwidth]{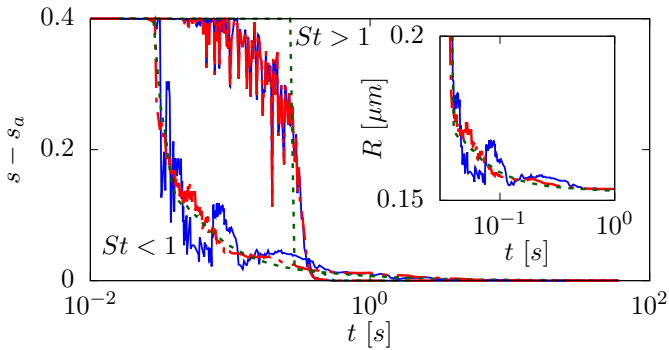}  
\caption{$s-s_a$ as a function of time experienced by two  representative droplets in the DNS (blue, continuous line), filtered DNS (red, long dashed line) and mean-field (green, dashed line) simulations.
The group of three curves close to the bottom-left corner of the figure corresponds to a `small droplet'  having  an initial radius of $0.6\times 10^{-6}\, \mathrm{m}$ and a Stokes number always smaller than 0.004 during the whole droplet evolution (referred to as $St<1$ in the figure).
The group of three curves in the upper part of the main figure  corresponds to a `large  droplet'  having  an initial radius of
$0.8\times 10^{-3}\, \mathrm{m}$  and  a Stokes number always larger than 3 during the whole droplet evolution (referred to  as $St>1$ in the figure). 
The inset shows the radius time evolution of the `small droplet'.}
\label{fig:S}
\end{figure}
This phenomenon can be clearly detected in Fig.~\ref{fig:S} where the temporal behavior of the supersaturation field along the Lagrangian trajectory
of a small airborne droplet is reported (group of lines denoted by $St<1$) together with the time evolution of the corresponding droplet radius (see the inset of Fig.~\ref{fig:S}). The time history with the fully resolved DNS (blue, continuous line) clearly shows the effect of the plateaux-and-cliffs structures on the evaporation process which is however absent for the larger sedimenting droplet 
(group of lines denoted by $St>1$). 
The fact that the radius closely follows the  temporal behavior of the supersaturation field (inset of Fig.~\ref{fig:S}) is the signature of a quasi-adiabatic picture for the evaporation process (i.e. the
process of radius adjustement due to evaporation is much faster than the corresponding variation of the supersaturation field).
It is worth noting that if one considers the smaller droplet evolving in coarse grained fields (long dashed line in red, where both velocity  and supersaturation have been coarse grained in space as discussed in Sec.~\ref{app:coarse-graining}),  the effect of the plateaux-and-cliffs structures on the evaporation process reduces, and eventually vanishes when the turbulent fields are replaced by their mean-field components (green dashed line).

\begin{figure}
\centering
\includegraphics[width=0.49\textwidth]{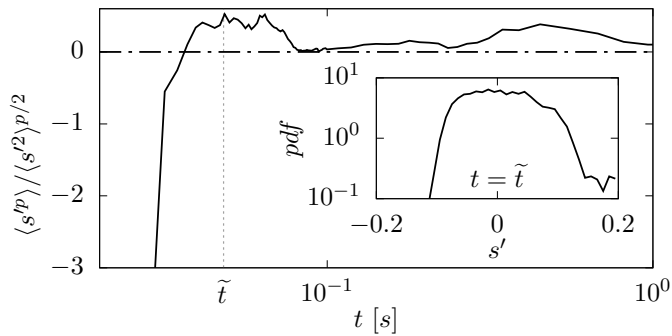}  
\caption{Skewness ($p=3$, continuous line)  and mean value ($p=1$, dot-dashed line) of the supersaturation turbulent fluctuation $s'$. Inset: the probability density function of $s'$ at the time 
$\widetilde{t}=0.05\ s$. Note the change of sign from negative to positive skewness passing from the jet to the puff phase.}
\label{fig:skew}
\end{figure}

Having shown that sufficiently small droplets correlate with the supersaturation field, let us now discuss the consequences on droplet motion.
For smaller droplets remaining for a sufficiently long time in regions where the supersaturation field is locally constant, with a value larger (smaller) than the mean, the evaporation takes place more slowly (rapidly) than what  would be for 
the same droplet experiencing smoother fluctuations as in the filtered DNS or in the mean-field approach.
The two effects, i.e. reduction vs increase in evaporation time, are however not symmetric as a consequence of a positive skewness observed in the probability density function of $s'$, the turbulent fluctuation of the supersaturation field. As shown in Fig.~\ref{fig:skew}, a positive skewness is accompanied by a zero-mean value of $s'$. The net result caused by turbulent fluctuations of the supersaturation field on the fate of small droplets is thus to increase their evaporation time.\\
Evidence of positive skewness has been reported for scalar concentration emitted by point sources within atmospheric turbulent flows~\cite{Nironi2015}.

Let us now quantify the delay caused by turbulence in the evaporation process by comparing, for an observation time of $60 \,\mathrm{s}$, the time it takes for each airborne droplet to shrink to their final equilibrium radius. Let us denote those typical evaporation times as $\tau_{evap}$.
All droplets which sedimented within the observation time of $60\,\mathrm{s}$
were not included in this analysis. The sole airborne droplets were selected here, thus automatically satisfying the requirement of having a sufficiently small radius.\\
\begin{figure}
\centering
\includegraphics[width=0.49\textwidth]{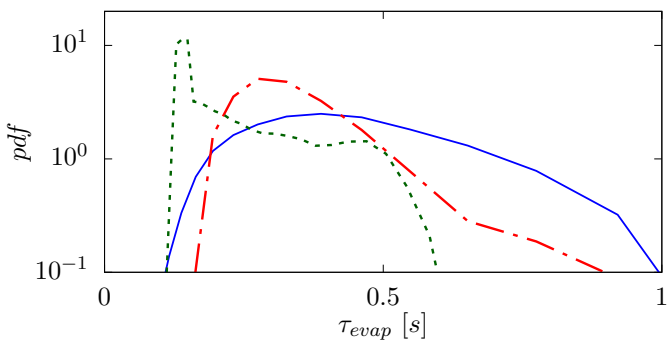}  
\caption{Probability density function of the time for each airborne droplet to shrink to its final equilibrium radius  for the DNS (blue, continuous line),
  filtered DNS (red, long dashed line) and mean-field (green, dashed line) simulations. Only airborne particles in the observation time of $60\,s$ are considered.}
\label{fig:evap}
\end{figure}
\begin{table}
\centering
\setlength{\tabcolsep}{5pt}
\begin{tabular}{c|ccc}
Simulation type	& DNS	& Filtered DNS	& Mean-field	\\
\hline
$\langle \tau_{evap}\rangle\, [s]$ 	& $0.4$	& $0.3$	& $0.2$	\\
\end{tabular}
\caption{Droplet mean evaporation times calculated from the probability density functions of Fig.~\ref{fig:evap}.}
\label{tab:evap}
\end{table}
The results are presented in Fig.~\ref{fig:evap} where the probability density functions of  $\tau_{evap}$ are reported both for the fully resolved case and for the evolution with the sole mean fields (of both the carrying flow and the supersaturation field) and with the filtered DNS. The corresponding mean evaporation times are reported in Tab.~\ref{tab:evap}.
The role of turbulence clearly emerges, both causing delay of the evaporation process and broader probability density functions, the fingerprint of fluctuations.
\begin{figure}
\centering
\includegraphics[width=0.49\textwidth]{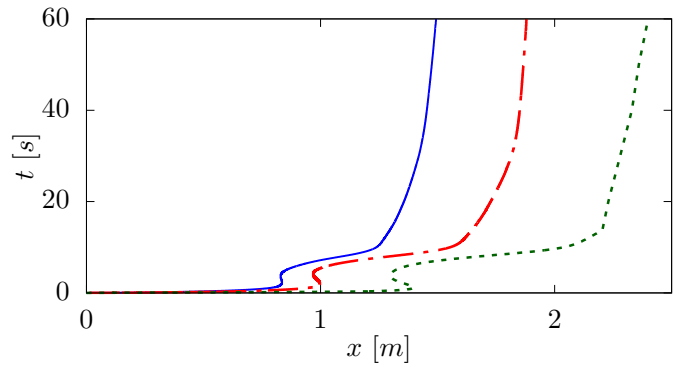}  
\caption{The streamwise coordinate, $x(t)$, of the center of mass of the cloud of airborne droplets.
  Blue, continuous line: DNS simulation;  red, long dashed line: filtered DNS simulation;  green, dashed line: mean-field simulation.}
\label{fig:sedSmall}
\end{figure}

Importantly, the observed delay in the evaporation significantly affects droplet motion. This is depicted in Fig.~\ref{fig:sedSmall} where we report
  the streamwise coordinate of the center of mass
of the cloud of airborne droplets, $x(t)$,
as a function of time. Shown in this figure are the fully resolved DNS, the filtered DNS, and the mean-field approach.
In the two cases where turbulent fluctuations are either coarse grained or entirely neglected, droplets travel further than in the fully resolved DNS. 
This is the fingerprint of the reduced inertia of the droplets evolving in the filtered fields. In the initial stage of their evolution, these droplets are indeed spuriously lighter than the droplets evolving in the fully-resolved DNS. Being lighter, they are carried more efficiently by the underlying rapidly accelerating flow thus reaching longer distances before touching the floor.
Note also that all the curves show a pronounced S-shaped kink which reflects the rapid evaporation of relatively large droplets exiting from the puff, resulting in a sudden reduction of the total mass of the droplet cloud.

\begin{figure}[t!]
\centering
\includegraphics[width=0.49\textwidth]{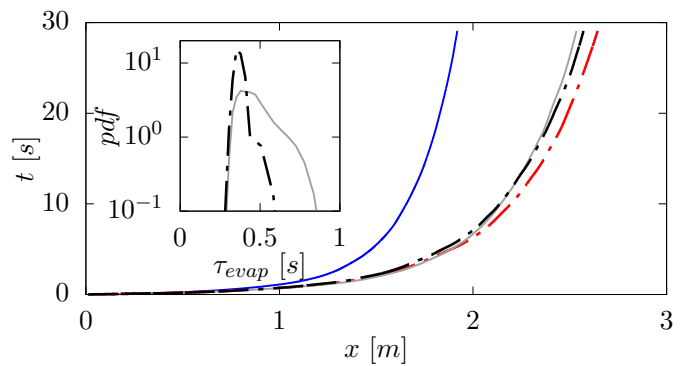}  
\caption{The streamwise coordinate, $x(t)$, of the center of mass of the cloud of airborne droplets.
  Results refer to the simulations for the monodisperse droplets of initial radius $R_i(0)=5\, \mathrm{\mu m}$ with and without inertia in the droplet ruling equations.
  Main frame: inertia causes differences in droplet trajectories.
 DNS with (without) inertia  are represented by the
  continuous blue (gray)  curve; filtered DNS  with (without) inertia by  the long-dashed red (black) curve. Inset: turbulence causes the observed broadening of evaporation times.  The probability density function of the evaporation time $\tau_{evap}$ without inertia for the DNS simulation (continuous gray curve) and for the  filtered DNS simulation (long-dashed black curve).}
\label{fig:mono}
\end{figure}

In order to ascertain whether the observed delay of trajectories of small droplets is a genuine effect 
caused by the interplay between turbulence and inertia, a subset of idealized simulations have been performed where monodisperse droplets of $R_i(0)=5\, \mathrm{\mu m}$ are considered, with and without inertia (i.e.\ simply switching on/off inertia in the ruling equations (\ref{eq:max1}) and (\ref{eq:max2})). This size is close to the peak of the droplet size distribution we have used in the previous analysis~\cite{duguid1946size}, and corresponds to droplets that are neither too large to be insensitive to turbulence, nor too small to make the mass loss due to evaporation negligible.

The results are shown in Fig.~\ref{fig:mono}. Both in the presence and in the absence of droplet inertia we found  the turbulence-induced  broadening of the probability density functions of the evaporation time. This is shown in the inset of Fig.\ \ref{fig:mono} for the simulations without inertia.
Filtering the turbulence fluctuations (long-dashed black curve in the inset) reduces the broadening as observed for the polydisperse case with inertia.
It is now worth remarking  that the  observed difference between the  mean evaporation time
measured from the DNS and the one measured from the filtered DNS
does not produce any relevant effect on the droplet motion when inertia is switched off in the droplet ruling equations. The similarity in the main frame between the continuous gray curve and the black long-dashed curve confirms this fact.
Switching-on inertia, the effect of the  delayed evaporation in the DNS case
becomes apparent (see  in the main frame the differences between the 
continuous blue curve and the red long-dashed curve).
Fig.~\ref{fig:mono} confirms that turbulence is the root cause of the broadening of evaporation times, whereas inertia causes differences in the trajectories.

\section{Conclusions}
\label{sec:conclusions}

In this work, we investigated the physical mechanisms involved in violent expiratory events such as coughing and sneezing, focusing on the evaporation and consequent airborne spread of small exhaled droplets. To this aim, we conducted a series of DNS experiments at unprecedented resolutions of the airflow associated with human cough~\cite{gupta2009flow} in order to fully resolve the turbulent fluctuations both in time and space. Droplet dynamics are evolved by means of a Lagrangian model including the evolution of droplet radius to properly describe the droplet evaporation process. 
Selecting a representative initial distribution of droplet sizes from current literature~\cite{duguid1946size}, we track each single droplet in time. We distinguish between larger droplets which settle on the ground ballistically and the smaller droplets which remain trapped in the turbulent puff. 

For such airborne droplets, we found that turbulence plays a crucial role in determining their evaporation time.
To demonstrate this result, we perfomed the same numerical experiments using two different coarse-graining techniques, i.e. filtered DNS and mean-field simulations. Compared to the DNS results, we find that coarse graining leads to underestimating droplet evaporation time up to $100\%$. 
Correspondingly, we find that DNS are crucial to accurately describe the inertial effects in droplet trajectory and ultimately predict their flight time and final reach. Importantly, the heated debate on social distancing rules depend crucially on these observables.

Do the same conclusions drawn here apply for sneeze expulsions? Sneezing  differs from coughing mainly for the larger Reynolds numbers and for the larger number of exhaled droplets. According to  \citet{duguid1946size} up to a million droplets may be emitted in a sneeze, compared to few thousand typical for cough. Thus droplets in a cough are far from one another, but this may not be the case for droplets emitted in a sneeze. In a sneeze, droplets may affect one another and well documented clustering effects may further delay evaporation~\cite{derivas2016dense,villermaux2017fine}. Further work is needed to clarify the potential role of clustering in delaying evaporation during violent human expulsions.
\\[15pt]



\noindent {\em Acknowledgements} A.M. thanks the financial support from the Compagnia di San Paolo, project MINIERA no. I34I20000380007. 
M.E.R., S.O. and A.M. acknowledge the computational time provided by HPCI on the Oakbridge-CX cluster in the Information Technology Center, The University of Tokyo, under the grant hp200157 of the ``HPCI Urgent Call for Fighting against COVID-19'' and the computer time provided by the Scientific Computing section of Research Support Division at OIST.
Useful discussions with G. Seminara, B. Carli, G. Forni, S. Fuzzi and A. Rinaldo are warmly acknowledged.

\appendix

\begin{table*}
\caption{Physical/chemical properties assumed in the present study.}
\label{tab:properties}
\begin{ruledtabular}
\begin{tabular}{lcc}
  Mean ambient temperature & $T$ & $25\,^{\circ}\mathrm{C}$\\
  Crystallization (or efflorescence) RH & $\mathrm{CRH}$ & $45\%$\\
  Deliquescence RH & $\mathrm{DRH}$ & $75\%$\\
  Quiescent ambient RH & $\mathrm{RH}_a$ & $60\%$\\
  Density of liquid water & $\rho_w$ & $9.97\times 10^2\, \mathrm{kg/m^3}$\\
  Density of soluble aerosol part (NaCl) & $\rho_s$ & $2.2\times 10^3\, \mathrm{kg/m^3}$\\
  Density of insoluble aerosol part (mucus) & $\rho_u$ & $1.5\times 10^3\, \mathrm{kg/m^3}$\\
  Mass fraction of soluble material (NaCl) w.r.t. the total dry nucleus & $\epsilon_m$ & $0.75$\\
  Mass fraction of dry nucleus w.r.t. the total droplet & ${\cal C}$ & $1 \, \%$\\
  Specific gas constant of water vapor & $R_v$ & $4.6\times 10^2\, \mathrm{J/(kg\, K)}$\\
  Diffusivity of water vapor & $D_v$ & $2.5\times 10^{-5}\, \mathrm{m^2/s}$\\
  Density of air & $\rho_a$ & $1.18\, \mathrm{kg/m^3}$\\
  Kinematic viscosity of air & $\nu$ & $1.8\times 10^{-5}\, \mathrm{m^2/s}$\\
  Heat conductivity of dry air & $k_a$ & $2.6\times 10^{-2}\, \mathrm{W/K\, m}$\\
Latent heat for evaporation of liquid water & $L_w$ & $2.3\times 10^6\, \mathrm{J/kg}$ \\
Saturation vapor pressure & $e_\mathit{sat}$ & $0.616\, \mathrm{kPa}$\\
Droplet condensational growth rate & $C_R$ & $1.5  \times 10^{-10} \,\mathrm{m^2/s}$\\
Surface tension between moist air and salty water & $\sigma$ & $7.6\times 10^{-2}\, \mathrm{J/m^2}$\\
Molar mass of NaCl & $M_s$ & $5.9\times 10^{-2}\, \mathrm{kg/mol}$\\
Molar mass of water & $M_w$ & $1.8\times 10^{-2}\, \mathrm{kg/mol}$\\
\end{tabular}
\end{ruledtabular}
\end{table*}

\section{Physical/chemical properties of cough}
\label{app:properties}

The complete list of physical and chemical parameters appearing in our model, along with their baseline values adopted in this investigation, is presented in Table~\ref{tab:properties}. Some of these quantities are deduced by other parameters. Specifically, the saturation vapor pressure above flat water surface at temperature $T$ (where $T$ is in degrees Celsius) is obtained using the Magnus-Tetens approximation~\cite{monteith2013principles}
\begin{widetext}
\begin{equation}
e_\mathit{sat}= 6.1078\times 10^2\,e^{(17.27\, T/(T+237.3))} \, \mathrm{Pa}
\end{equation}
and the droplet condensational growth rate is given by
\begin{equation}
C_R=\left[\frac{\rho_w\, R_v\, (273.15+T)}{e_{sat}\,D_v}+\frac{\rho_w\,L_w^2}{k_a\,R_v\,(273.15+T)^2}-\frac{\rho_w\,L_w}{k_a(273.15+T)}\right ]^{-1}.
\end{equation}
\end{widetext}

The expressions of the coefficients $A$ and $B$ appearing in
Eq.\ (7) follow from~\citet{pruppacher1997microphysics} (p.\ 176):
\begin{equation}
A=\frac{2 \sigma}{R_v (T+273.15) \rho_w}, 
\end{equation}
\begin{equation}
B=\frac{n_s \Phi_s \epsilon_v M_w \rho_s}{M_s \rho_w},
\end{equation}
where $n_s=2$ is the total number of ions into which a salt molecule dissociates, $\Phi_s=1.2$ is the practical osmotic coefficient of the salt in solution~\cite{liu2017short} and $\epsilon_v= \epsilon_m(\rho_N/\rho_s)$ is the volume fraction of dry nucleus with respect to the total droplet.

To complete the description, some useful relations can be easily derived from the quantities specified in Table~\ref{tab:properties}.
First, assuming that the dry nucleus of droplets is composed by a soluble phase (NaCl) and an insoluble phase (mucus) and that the typical value of the mass fraction of the former is known, the overall density of the dry nucleus can be expressed as 
\begin{equation}
\rho_N  = \frac{\rho_u}{1-\epsilon_m[1-(\rho_u/\rho_s)]}= 1.97\times 10^3\, \mathrm{[kg/m^3]}.
\end{equation}
Similarly, the density of the entire $i$-th droplet turns out to be
\begin{equation}
\rho_{D\,i}  = \rho_w + (\rho_N-\rho_w)\left (\frac{r_{N\,i}}{R_i(t)}\right )^3,
\end{equation}
where the radius of the (dry) solid part of the droplet when NaCl is totally crystallized (i.e. below CRH) is given by
\begin{equation}
r_{N\,i}  = R_i(0)\left (\frac{{\cal C}\;\rho_w}{{\cal C}\;\rho_w +\rho_N(1-{\cal C})}\right )^{1/3}.
\end{equation}

\section{Details on the Lagrangian model for the droplet dynamics}
\label{app:droplet}
We take as a model for the dynamics of an individual droplet the stochastic differential equations with additive noise~\cite{gatignol1983faxen,maxey1983equation}:
\begin{equation}
  \dot{\bm{X}}=\bm{U}(t)+\sqrt{2 \kappa_1}\bm{\eta} (t)
  \end{equation}
\begin{equation}
  \dot{\bm{U}}=\frac{\bm{u}(\bm{X}(t),t)-\bm{U}(t)}{\tau}+\sqrt{2 \kappa_2}\bm{\mu} (t)
\end{equation}
where the same notation of Sec.~\ref{sec:method} is used having dropped the droplet index.
In the above equations,  vectors
$\bm{\eta}(t)$ and $\bm{\mu}(t)$ denote independent white noises with Brownian diffusivity constants $\kappa_1$ and $\kappa_2$. The reason for considering a non-vanishing Brownian force acting on the position process is twofold and detailed in Ref.~\cite{afonso2018eddy}.
In the limit of tracer particles, i.e.~$\tau\to 0$ in the above equations,
the quantity $\kappa_1+\kappa_2$ can be immediately identified with the
(water vapor) diffusivity $D_v$. Namely, $D_v=\kappa_1 + \kappa_2$.
Because of the fact that the only accessible value here is the one of $D_v$,
we opted for the simplest choice $D_v=\kappa_1$ which  guarantees
a viscous regularization of the large-scale transport~\cite{afonso2018eddy}.

\section{Numerical method}
\label{app:num-meth}

\begin{figure*}
    \centering
    \includegraphics[width=0.45\textwidth]{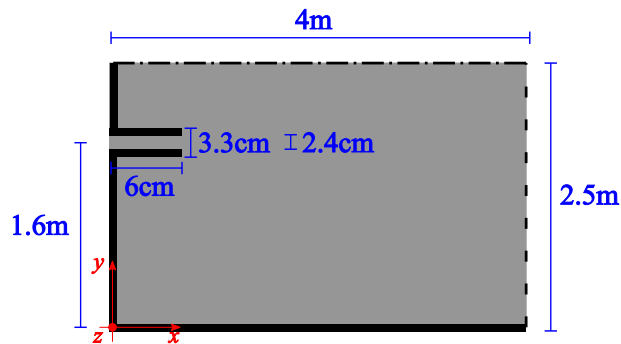} 
    \caption{Sideview sketch of the domain used in our DNS (note that the figure is not to scale).}
    \label{fig:sketch-size}
\end{figure*}

\begin{figure*}
    \centering
    \includegraphics[width=0.45\textwidth]{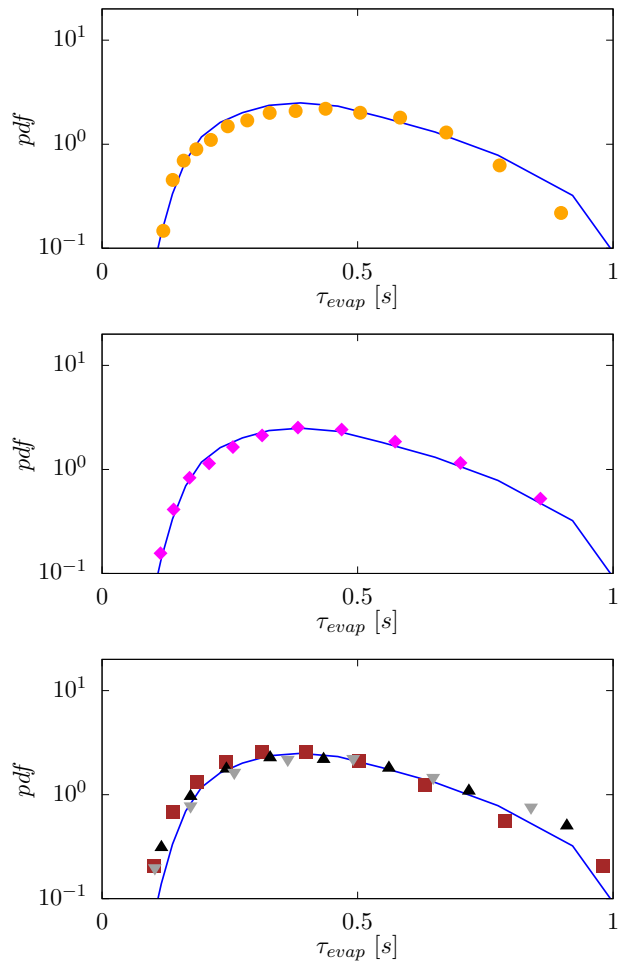}
    \caption{(top) Grid convergence analysis: probability density function of the particle evaporation time computed in the baseline case with spacing $\Delta x=3.5\,\mathrm{mm}$ (blue curve) and $\Delta x=1.75\,\mathrm{mm}$ (orange symbols).
    (middle) Convergence of the statistics: probability density function of the particle evaporation time computed in the baseline case with $N\approx5000$ (blue line) and with half of them (magenta symbols).
    (bottom) Independency of the results from the initial condition: probability density function of the particle evaporation time computed in the baseline case (blue line) and when droplets are emitted with a time delay of $0.07\,\mathrm{s}$ (brown), $0.14\,\mathrm{s}$ (black) and $0.21\,\mathrm{s}$ (gray).}
    \label{fig:evapTimeGrid}
\end{figure*}

In this section, we supply additional information on the computational framework that is used to investigate the problem.
The fluid flow equations [Eqs.~\eqref{eq:NS} and~\eqref{eq-supersat}] are solved within a domain box of length $L_x=4\,\mathrm{m}$, height $L_y=2.5\,\mathrm{m}$ and width $L_z=1.25\,\mathrm{m}$, as depicted in Fig.~\ref{fig:sketch-size}. The fluid is initially assumed at rest, i.e. $\bm{u}(\bm{x},0)=\bm{0}$. Air is thus injected through a circular pipe, placed at $z=1.6\,\mathrm{m}$ above the floor, of length $l=6\,\mathrm{cm}$ and internal diameter $d=2.4\,\mathrm{cm}$ as an essential model of a human mouth. We use the time-varying velocity profile proposed by~\citet{gupta2009flow} (shown in Fig.~\ref{fig:inletVel}) to reproduce the cough-associated airflow.
The no-slip condition applies at the bottom, i.e. $y=0$ and left wall, i.e. $z=0$ (solid lines in Fig.~\ref{fig:inletVel}). At the top ($y=L_y$, dot-dashed), we prescribe the free-slip condition.
For the supersaturation field $s$, at $t=0$ we have $s(\bm{x},0)=s_a=\mathrm{RH}_a-1$ everywhere in the domain.
The inlet flow exiting from the mouth is assumed to be saturated air, i.e. $s=0$. The Dirichlet condition $s=s_a$ is thus used at the bottom, left and top boundaries.
For both the velocity and supersaturation field, we impose a convective outlet boundary condition at the right boundary $x=L_x$ (dashed line). Finally, periodic boundary conditions apply at the side walls, i.e. $z=0$ and $z=L_z$.

In our simulations, the domain is discretized with uniform spacing $\Delta x = 3.5 \, \mathrm{mm}$ in all directions, resulting in a total number of $N \approx 0.3$ billion grid points.
Results are validated against the theoretical prediction for a turbulent puff (see Fig.~2). Moreover, we assessed the convergence with respect to the grid resolution, as it is shown in Fig.~\ref{fig:evapTimeGrid} (top) where we compare the probability density function of the particle evaporation time using the adopted grid setting with that obtained by doubling the spatial resolution. From the figure we can clearly observe that only minor differences occur, thus confirming the reliability of the chosen grid resolution. 

\sloppy The results discussed in the text are statistically significant. We varied this by halving the numerical sample (Fig.~\ref{fig:evapTimeGrid} (middle)) and by varying the release time of the droplets (Fig.~\ref{fig:evapTimeGrid} (bottom)), thus resulting in different dynamics due to the chaotic nature of the flow; for both tests the figure shows no appreciable differences.


%

\end{document}